\documentclass[runningheads]{llncs}
\usepackage[T1]{fontenc}
\usepackage{graphicx}
\usepackage{amsmath}
\usepackage{amssymb}
\begin{document}

\title{Conformal Performance Range Prediction for Segmentation Output Quality Control}
\titlerunning{Conformal Performance Range Prediction for Output Quality Control}
%

\author{Anna M. Wundram\inst{1} \and
Paul Fischer\inst{1,5} \and Michael Mühlebach\inst{2} \and 
Lisa M. Koch\inst{3,4} \and Christian F. Baumgartner \inst{1,5}}

\authorrunning{A. M. Wundram et al.}
%
\institute{Cluster of Excellence -- ML for Science, University of Tübingen, Germany\\ \email{anna.wundram@student.uni-tuebingen.de, paul.fischer@uni-tuebingen.de}
\and
Max Planck Institute for Intelligent Systems, Tübingen, Germany
\email{michael.muehlebach@tuebingen.mpg.de}
\and
Department of Diabetes, Endocrinology, Nutritional Medicine and Metabolism UDEM, Inselspital, Bern University Hospital, University of Bern, Switzerland
\and
Hertie Institute for AI in Brain Health,  University of Tübingen, Germany
\email{lisa.koch@unibe.ch}
\and
Faculty of Health Sciences and Medicine, University of Lucerne, Switzerland
\email{christian.baumgartner@unilu.ch}}
\maketitle              

\begin{abstract}
\setcounter{footnote}{0} 
Recent works have introduced methods to estimate segmentation performance without ground truth, relying solely on neural network softmax outputs. These techniques hold potential for intuitive output quality control.
However, such performance estimates rely on calibrated softmax outputs, which is often not the case in modern neural networks. Moreover, the estimates do not take into account inherent uncertainty in segmentation tasks. These limitations may render precise performance predictions unattainable, restricting the practical applicability of performance estimation methods.  
To address these challenges, we develop a novel approach for predicting performance \textit{ranges} with statistical guarantees of containing the ground truth with a user specified probability. Our method leverages sampling-based segmentation uncertainty estimation to derive heuristic performance ranges, and applies split conformal prediction to transform these estimates into rigorous prediction ranges that meet the desired guarantees. 
We demonstrate our approach on the FIVES retinal vessel segmentation dataset and compare five commonly used sampling-based uncertainty estimation techniques. Our results show that it is possible to achieve the desired coverage with small prediction ranges, highlighting the potential of performance range prediction as a valuable tool for output quality control\footnote{Code available at https://github.com/annawundram/PerformanceRangePrediction}.
\end{abstract}
\section{Introduction}

Image segmentation is a crucial step for various medical tasks such as disease detection, treatment planning or anatomical studies~\cite{azad2022medical}. In ophthalmology, segmenting the retinal vessels in fundus photography images provides insights into clinical conditions such as Glaucoma or Diabetic Retinopathy~\cite{jin2022fives}. However, manual segmentation of the retinal vasculature is prohibitively time-consuming, taking up to five hours per image~\cite{jin2022fives}. In recent years, machine learning models have shown excellent performance on many segmentation problems~\cite{isensee2021nnu,bernard2018deep}. Despite this, machine learning methods can fail unexpectedly, and even the best algorithms have limited performance on images that are inherently challenging to segment. As manual verification of all algorithmic outputs is in itself time-consuming and sometimes infeasible, developing strategies for automatically ensuring the quality of segmentation outputs is becoming increasingly important.

\begin{figure}[t]
    \centering
    \includegraphics[width=\textwidth]{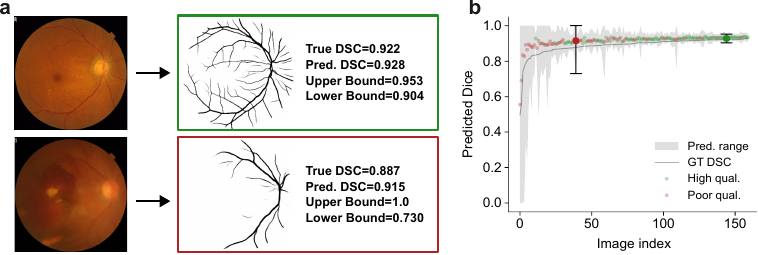}
    \caption{\textbf{Overview.} (a) Given a fundus image, we predict a vessel segmentation, the expected Dice-Sørensen Coefficient (DSC), as well as upper and lower bounds for the expected DSC. (b) Conformal prediction allows us to set the performance range such that at most $\alpha=10\%$ percent of the test cases have a DSC outside the predicted interval. We show a confident case with low DSC prediction uncertainty (green), as well as a case with high DSC prediction uncertainty due to poor image quality (red).}   
    \label{fig:overview}
\end{figure}

Ensuring segmentation quality can be approached through input or output quality control. Input quality control aims to automatically identify images that are likely to be poorly segmented by the model. Commonly used strategies include automatic prediction of image quality~\cite{zhang2016automated} or out-of-distribution (OOD) detection~\cite{liu2023residual,cho2023training}. However, input quality control methods may misjudge the algorithm's actual performance on a specific image. For instance, as demonstrated in this work, an image may be within the support of the training distribution yet difficult for the model to segment, or it may be of poor quality but still easy to segment (see Fig.~\ref{fig:overview}). Such cases might either go undetected by input quality control, or could be flagged unnecessarily.

Output quality control methods instead aim to directly verify that the output of a model is of sufficient quality. Most commonly, output uncertainty is used as a  proxy for quality (e.g.~\cite{lin2022novel,ng2018estimating,puyol2020automated}). However, these approaches require choosing a heuristic threshold for acceptable quality. To address this, a number of methods directly estimate the expected performance on previously unseen data points. This allows setting more intuitive performance thresholds such as ``a Dice-Sørensen Coefficient (DSC) of at least 0.8''. In one of the first works on medical segmentation performance prediction, \cite{kohlberger2012evaluating} trained a regressor to directly predict segmentation accuracy.
Later, Valindra et al.~\cite{valindria2017reverse} introduced Reverse Classification Accuracy, which trains a reverse segmentation model on the test image and its segmentation output, then applies it to reference images to estimate the DSC. However, this approach requires training a second model.

Recent studies have demonstrated that segmentation performance measures can be predicted from softmax outputs alone~\cite{williams2021automatic,fournel2021medical,sunoqrot2020quality,herrera2020framework,robert2018real_time,hahn2021ensemble,koehler2024efficiently,li2022estimating}. Assuming perfect calibration, the softmax outputs indeed describe the probability of a pixel having a certain label. As detailed in Sec. \ref{sec:DSCprediction}, this interpretation allows for the computation of various performance measures, including the DSC~\cite{li2022estimating,koehler2024efficiently}.
However, the quality of this performance estimator heavily relies on proper calibration~\cite{koehler2024efficiently}. Unfortunately, modern neural networks frequently exhibit poor calibration~\cite{guo2017calibration}. Although post-hoc methods like temperature scaling can enhance calibration~\cite{koehler2024efficiently,guo2017calibration}, these solutions are often insufficient for achieving the desired level of reliability. Moreover, inherent segmentation uncertainty arising from low image quality or other factors, may further limit precise performance estimation. As a result, the performance estimates obtained through these methods lack guarantees, raising concerns about their usefulness for quality control.

In this work, we propose to predict performance \textit{ranges} instead of point estimates. Our approach offers statistical guarantees that the true performance of any test image falls within the predicted range with a certain probability. We achieve this by estimating a heuristic performance range using sampling-based uncertainty quantification methods. We then apply split conformal prediction~\cite{vovk2005algorithmic,angelopoulos2021gentle} to conformalize prediction ranges extracted from those uncertainty estimates. We demonstrate the effectiveness of our approach on the challenging problem of retinal vessel segmentation in fundus images.

\section{Methods}

Given an input image $x$ our goal is to predict a segmentation $\hat{s}$ along with a performance estimate $\hat{y}$ predicting the model's segmentation performance (e.g. the DSC) for that image. We additionally predict a performance range $[\hat{y}_l, \hat{y}_u]$ with a statistical guarantee that the true DSC, i.e. $y=\textrm{DSC}(\hat{s}, s)$, is contained in this interval with a user specified probability of $1-\alpha$. Note that while we focus on the DSC, alternative measures can be easily investigated in our framework. 

In the following, we first review how a DSC performance estimate can be derived from the softmax outputs of a segmentation model (Sec.\,\ref{sec:DSCprediction}). Next, we introduce our method for obtaining heuristic performance ranges using sampling-based segmentation uncertainty estimation approaches (Sec.\,\ref{sec:perf-bounds-uncertainty}). Lastly, we describe our strategy for converting heuristic performance ranges, into performance ranges with statistical guarantees using split conformal prediction (Sec.\,\ref{sec:principledbounds}). 

\subsection{Background: Estimating the DSC from softmax outputs}
\label{sec:DSCprediction}

Given a calibrated segmentation model, the softmax output $p_i$ for each pixel $i$ can be interpreted as the probability of that pixel being of the predicted class. In the binary case, summing the positively predicted (i.e. $p_i > 0.5$) foreground probabilities for all pixels yields the expected number of true positives (TP) in the image. Following a similar reasoning the expected number of false positives (FP) and false negatives (FN) can be calculated \cite{li2022estimating}:
\begin{equation}
\label{eq:confusion-terms}
    \textrm{TP} = \sum^n_{i = 1} \mathbf{1}_{[p_i > 0.5]}p_i;~
    \textrm{FP} = \sum^n_{i = 1} \mathbf{1}_{[p_i > 0.5]} - \textrm{TP};~
    \textrm{FN} = \sum^n_{i = 1} \mathbf{1}_{[p_i < 0.5]}p_i \ \text{.}
\end{equation}
Here, $\mathbf{1}_{[\cdot]}$ denotes the indicator function and $0.5$ is a heuristically chosen decision threshold.
The DSC is defined in terms of those quantities as follows
\begin{equation}
\label{eq:definition-dsc}
    \textrm{DSC} = \dfrac{2\textrm{TP}}{2\textrm{TP} + \textrm{FP} + \textrm{FN}}\ \text{.} 
\end{equation}
Thus, for a given test image $x$, a DSC performance estimate $\hat{y}$ can be calculated using the estimators for $\textrm{TP}$, $\textrm{FP}$, and $\textrm{FN}$ (Eq.~\ref{eq:confusion-terms}), and plugging them into Eq.~\ref{eq:definition-dsc}. 

In practice neural networks are never perfectly calibrated, resulting in over- or underestimation of the true DSC. Instead of relying on the possibly faulty performance prediction $\hat{y}$, in the following, we show how to obtain bounds that contain the true performance with high probability.

\subsection{Heuristic performance bounds from segmentation uncertainty}
\label{sec:perf-bounds-uncertainty}

To obtain heuristic lower and upper performance bounds $\bar{y}_l$ and $\bar{y}_u$, we rely on probabilistic segmentation techniques that are capable of producing samples from the distribution $p(s|x)$. Given an input image $x$ these techniques allow us to sample $N$ plausible segmentation samples $\hat{s}_n$. A performance estimate $\hat{y}_n$ can be obtained from each segmentation sample $\hat{s}_n$ as described in Sec.\,\ref{sec:DSCprediction}. 

The samples $\hat{y}_n$ characterize the distribution $p(\hat{y}|x) = \int p(s|x)p(\hat{y}|s)ds$. From these samples, we can calculate an estimator for the standard deviation $\sigma$.
We can then define our heuristic upper and lower bounds for an input image $x$ as: 
\begin{equation}
\label{eq:heuristic-ranges}
    \bar{y}_l(x) = \hat{y}(x) - \sigma(\hat{y}(x)); ~ ~ ~ ~ \bar{y}_u(x) = \hat{y}(x) + \sigma(\hat{y}(x)) ~\ \text{.}
\end{equation}

We compare five commonly used probabilistic segmentation techniques to obtain segmentation samples for performance range prediction: 
\begin{itemize}
    \item The \textbf{probabilistic U-Net}~\cite{kohl2018probabilistic} is a combination of the conditional VAE~\cite{sohn2015learning} approach with a U-Net architecture. This formulation allows to sample an infinite number of segmentation samples consistent with the input image $x$. 
    \item \textbf{PHiSeg}~\cite{phiseg} extends the probabilistic U-Net by a hierarchical latent space and was shown to provide closer approximations of $p(\*s|\*x)$. 
    \item \textbf{Test-time augmentation (TTA)}~\cite{ayhan2018testtime} augments the test image $N$ times to obtain $N$ segmentation samples $s_n$. The deviations between the samples $s_n$ have been shown to be indicative of segmentation uncertainty. Following \cite{ayhan2018testtime}, we used the following eight types of augmentations: brightness, hue, saturation, contrast, vertical and horizontal flip, Gaussian blur.
    \item \textbf{Ensembles}~\cite{lakshminarayanan2017simple} consist of $N$ independently trained segmentation models initialized with different random seeds. In order to improve calibration, temperature scaling~\cite{guo2017calibration} on the calibration set was applied to each individual network. Note that temperature scaling is not directly applicable to any of the other explored methods. 
    \item \textbf{Monte Carlo (MC) Dropout}~\cite{gal2016dropout} produces probabilistic segmentation samples by repeatedly predicting segmentations for the same image with dropout enabled. We use a dropout rate of 0.2 on the activation maps for training and testing. Dropout is applied to all layers except the final four segmentation layers.
\end{itemize}
We use the enhanced U-Net architecture introduced in~\cite{kohl2018probabilistic} as a base architecture for all methods except PHiSeg. While PHiSeg also uses the same U-Net encoder, it employs a unique decoder.
We use $N=100$ for Prob. U-Net, PHiSeg and MC Dropout, $N=20$ for TTA, and $N=10$ for Ensembles. For all approaches, a final segmentation $\hat{s}$ is obtained by averaging the samples. 
The probabilistic U-Net, PHiSeg, and TTA estimate aleatoric uncertainty, while Ensembles and MC Dropout estimate epistemic uncertainty (see \cite{kahl2024values} for definitions of aleatoric and epistemic).

\subsection{From heuristic to principled bounds using conformal prediction}
\label{sec:principledbounds}

We employ split conformal prediction~\cite{vovk2005algorithmic,angelopoulos2021gentle} to convert the heuristic bounds $\bar{y}_l, \bar{y}_u$ into principled bounds $\hat{y}_l, \hat{y}_u$. Specifically, we desire that the performance range $[\hat{y}_l, \hat{y}_u]$ includes the true DSC $y$ with a user set probability of at least $1 - \alpha$:
\begin{equation}
\label{eq:conformal-probability}
     \mathbb{P} \left(y \in [\hat{y}_l(x), \hat{y}_u(x)] \right) \geq 1 - \alpha \ \text{.}
\end{equation}
We set $\alpha = 0.1$ in all our experiments. 

For our performance range to statistically fulfill the above requirement, we adjust the heuristic bounds with a corrective factor $\hat{q}$:
\begin{equation}
\label{eq:range-adjustment}
    \hat{y}_l(x) = \hat{y}(x) - \hat{q}\sigma(\hat{y}(x)); ~ ~ ~ ~ \hat{y}_u(x) = \hat{y}(x) + \hat{q}\sigma(\hat{y}(x)) \ \text{.}
\end{equation}
This corrective factor $\hat{q}$ is determined using the split conformal procedure~\cite{vovk2005algorithmic,angelopoulos2021gentle}. We first define a \textit{score function} as 
\begin{equation}
    \label{eq:scorefunction}
    \mathcal{S}(x,y) = \dfrac{|y - \hat{y}(x)|}{\sigma(\hat{y}(x))} ~ \text{.}
\end{equation}
The adjusting factor $\hat{q}$ is then calculated as the $\lceil (1-\alpha)(M+1) \rceil / M$ quantile on the calibration set scores, where $M$ is the number of calibration samples.
As shown in~\cite{vovk2005algorithmic}, this results in the following guarantee for the test set, assuming the calibration set is representative of the test distribution:
\begin{equation}
\mathbb{P} [\mathcal{S}(x, y) \leq \hat{q}] \geq 1-\alpha \Rightarrow \mathbb{P}[|y - \hat{y}(x)| \leq \sigma(\hat{y}(x))\hat{q}] \geq 1 - \alpha ~ \text{,}
\end{equation}
thereby fulfilling our requirement in Eq.~\ref{eq:conformal-probability}. The right-hand side of this equation follows from the definition in Eq.~\ref{eq:scorefunction}. We finally clamp the prediction range to be within $[0,1]$ as DSC values outside this range are not possible. 

\subsection{Data and training}
We evaluated our method on the FIVES~\cite{jin2022fives} fundus dataset for retinal vessel segmentations. The dataset comprises 800 images and manual segmentations with an official split into 600 training and 200 test images. We further split each fold using a ratio of 80/20 to obtain train/validation as well as test/calibration sets. Following~\cite{koehler2024efficiently}, we preprocessed the data by applying contrast limited adaptive histogram equalization (CLAHE) and resized the images to $320 \times 320$ pixels. We used the provided image quality labels for model inspection and visualization. We considered an image to be low quality if at least one out of three quality issues (illumination and color distortion, blur and low contrast) were reported.

All segmentation models were trained with maximum number of epochs of 1000 on a NVIDIA GeForce RTX 2080 Ti with a batch size of four for all probabilistic models and a batch size of 16 for all U-Nets. Model selection was performed on the validation set using the DSC as metric.

\section{Experiments and results}

\subsection{Evaluation of segmentation performance and DSC prediction}

\begin{figure}[t]
    \centering
    \includegraphics[width=\textwidth]{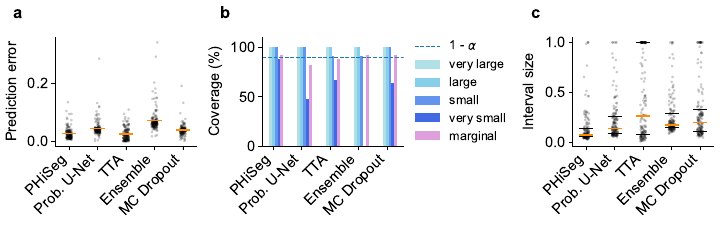}
    \caption{\textbf{Quantitative analysis.} (a) performance prediction absolute error, (b) marginal and conditional coverage for very small (0, 0.1], small (0.1, 2], large (0.2, 5], very large (0.5, 1] interval sizes, and (c) interval sizes for all investigated methods}  
    \label{fig:main-results}
\end{figure}

We first verified that all compared models performed adequately at the underlying segmentation task. We observed high mean test DSC for Prob. U-Net ($0.918$), MC Dropout ($0.918$) and Ensemble ($0.913$), with PHiSeg ($0.888$) performing slightly worse, though still acceptably, and in line with previous results on the FIVES dataset~\cite{koehler2024efficiently}. 
TTA ($0.811$) performed substantially worse than the rest of the evaluated techniques.

Next, we turned to the analysis of the performance prediction. PHiSeg and TTA were the most accurate at predicting the DSC, achieving mean absolute errors (MAE) of 0.027 and 0.025, respectively (see Fig.~\ref{fig:main-results}). Ensembles performed worst with a MAE of 0.072. This can be confirmed qualitatively by comparing the predictions to the ground truth line in Fig.~\ref{fig:case-coverage-plots}.

\subsection{Evaluation of coverage and interval sizes}

As we argue in this paper, a point estimate of the predicted performance is insufficient because poor calibration of the segmentation models lead to inaccurate scores, and because some cases carry inherent uncertainty in their performance prediction (e.g. due to low image quality).

\begin{figure}[t]
    \centering
    \includegraphics[width=\textwidth]{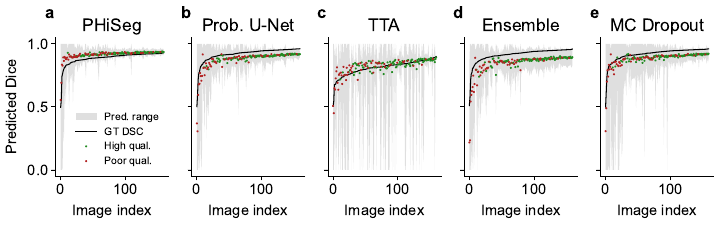}
    \caption{\textbf{Visualisation of performance ranges.} Performance predictions $\hat{y}$ (green/red), ground truth DSC scores $y$ (black), and performance ranges $[\hat{y}_l, \hat{y}_l]$ (gray) for all images in the test set. The images are sorted by ground-truth performance.}  
    \label{fig:case-coverage-plots}
\end{figure}

In our central evaluations, we therefore investigated the quality of the performance ranges obtained using our proposed method. We adopted the approach of~\cite{angelopoulos2022image} and used \emph{coverage} as our main evaluation criterion. In our case, coverage measures the proportion of images for which the actual DSC falls within the predicted performance range. Marginal coverage describes the coverage for a random test point. PHiSeg, Ensembles, and MC Dropout reach the specified marginal coverage, meaning that $\geq 90\%$ of all test ground truth DSC values lie in the predicted interval (Fig.~\ref{fig:main-results}b). We note that Prob. U-Net and TTA did not quite achieve marginal coverage. We hypothesize that this is due to a combination of poor uncertainty estimation and a slight violation of the exchangeability between calibration and test set assumption. The coverage of the studied segmentation models can also be analyzed qualitatively in Fig.~\ref{fig:case-coverage-plots} by verifying that the majority of ground truth DSC scores (black line) lie withing the gray prediction range. 

While split conformal prediction only guarantees marginal coverage, it is crucial for practical applications that coverage holds across different interval sizes. Therefore, in Fig.~\ref{fig:main-results}b, we also evaluated conditional coverage for four different interval sizes. This denotes the coverage for a test point belonging to a specific class (here: the interval size). All methods achieved the desired coverage for small (0.1, 0.2], large (0.2-0.5] and very large (0.5-1] intervals. PHiSeg achieved the best coverage for very small (0-0.1] interval sizes, but fell slightly short of the desired 90\%. Note that the bar for very small interval sizes is missing for Ensembles because no intervals of this size were predicted by the method. 

Assuming coverage is fulfilled, it is desirable to have interval sizes that are as small as possible. Overly large interval sizes resulting from poor DSC estimation or poor uncertainty estimation, may detract from the usefulness of the method in practice. It is therefore desirable to have interval sizes that are as small as possible. PHiSeg produced the tightest intervals (Fig.~\ref{fig:main-results}c), which can also be confirmed visually in Fig.~\ref{fig:case-coverage-plots} by inspecting the size of the gray prediction ranges.

Since the uncertainty in our task stems largely from irreducible ambiguities in the vessel segmentation, aleatoric uncertainty quantification methods should perform the best. Indeed, the top-performing approach, PHiSeg, falls into this cateogry. However, the overall picture is less clear, as the epistemic Ensemble and MC Dropout approaches perform similarly to the aleatoric Prob. U-Net. We concur with Kahl et al.~\cite{kahl2024values} that the distinction between aleatoric and epistemic uncertainty quantification methods is not always clear cut.

\subsection{Performance prediction analysis of low- vs. high-quality images}

To better understand the influence of image quality on the performance prediction, we colored all points in Fig.~\ref{fig:case-coverage-plots} by high-quality (green) and poor-quality (red) using the quality labels provided by the FIVES dataset. Firstly, as expected, we observed that most images with poor segmentation performance were of low quality. Secondly, overall performance prediction was worse for low-quality images compared to high-quality images. Thirdly, we observed that the size of the prediction performance range correlated with the ground truth DSC, indicating that harder-to-segment images also had higher uncertainty in their performance predictions.
The low-quality, low DSC images on the left side of the plots in Fig.~\ref{fig:case-coverage-plots} were typically characterized by large performance ranges. For the best-performing method, PHiSeg, although the performance estimation was poor in these cases, the intervals consistently contained the true DSC.
This illustrates that for these highly uncertain cases a single performance prediction is insufficient. It underscores that the statistically valid prediction performance ranges proposed here offer a promising approach for output quality control.

We note that all low-quality images used in this evaluation are in-distribution as the training set also contained similar low-quality images. This highlights the fact that OOD approaches would likely not be able to identify these cases where performance is low. An alternative strategy would be to train an image quality classifier to detect images with low-quality before feeding them to the model. However, there are also examples in Fig.~\ref{fig:case-coverage-plots} of low-quality images that achieve high DSC. An image quality classifier would falsely flag these images potentially resulting in unnecessary re-scans. Our proposed output quality control approach using performance ranges effectively addresses both these issues.

\section{Discussion and conclusion}

In this work, we demonstrated that performance prediction point estimates may be insufficient for robust quality control due to suboptimal calibration of neural networks, and high performance uncertainty in low-quality images. To address this problem, we developed a method that can compute performance \emph{ranges} with statistical guarantees for coverage, and compared five different sampling-based uncertainty quantification methods to estimate those range.

The aleatoric PHiSeg method produced the best performance predictions, as well as the performance ranges with the best coverage and tightest interval sizes. We conclude that this method is highly suitable for use in conformal performance prediction. 

A limitation of our work is that it is only applicable under the assumption of exchangeability of the test and calibration sets. The method is thus not directly applicable to the OOD setting. In future work, we will pursue an extension of our approach that takes advantage of novel research directions in conformal predictions under domain shifts~\cite{barber2023conformal}. 

\begin{credits}
\subsubsection{\ackname} This work was supported by the German Science Foundation (BE5601/8-1) and the Excellence Cluster 2064 ``Machine Learning --- New Perspectives for Science'', project number 390727645) and the Hertie Foundation. The authors thank the International Max Planck Research School for Intelligent Systems (IMPRS-IS) for supporting Paul Fischer.

\subsubsection{\discintname}
The authors have no competing interests to declare
that are relevant to the content of this article. 
\end{credits}
%
%
%
 \bibliographystyle{splncs04}
 \bibliography{bibliography}

\clearpage
\appendix
\section*{Supplementary Materials}

\renewcommand{\thesubsection}{\Alph{subsection}}

\subsection{Qualitative examples}
\begin{figure}
    \centering
    \includegraphics[width=0.9\textwidth]{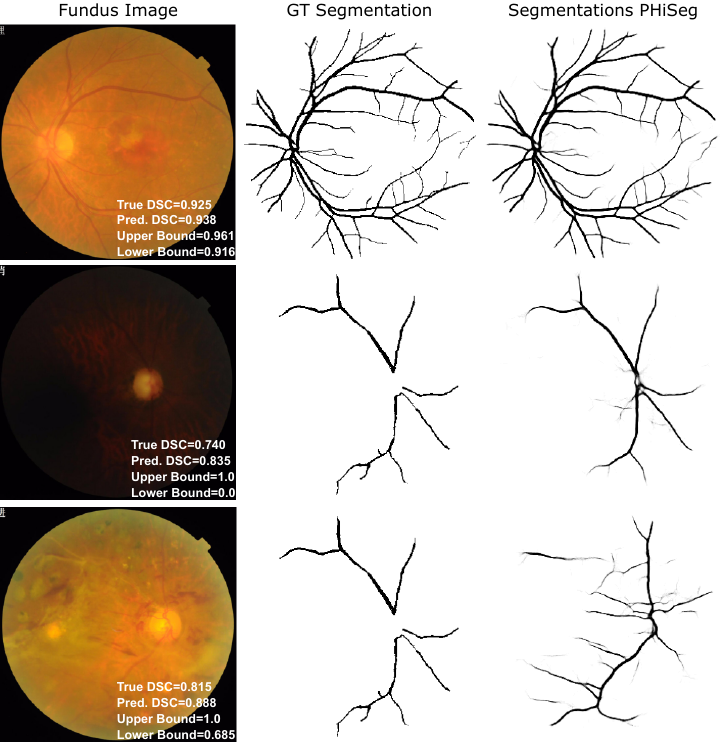}
    \caption{PHiSeg segmentations and performance predictions for three examples. Top: An example with good segmentation performance. Middle: An example with poor image quality and poor segmentation performance. Bottom: An example with good image quality but poor segmentation performance due to visible pathologies stemming from advanced diabetic retinopathy.}  
    \label{fig:exmaples}
\end{figure}
\clearpage
\subsection{Conditional coverage grouped by quality}
\begin{figure}
    \centering
    \includegraphics[width=0.49\textwidth]{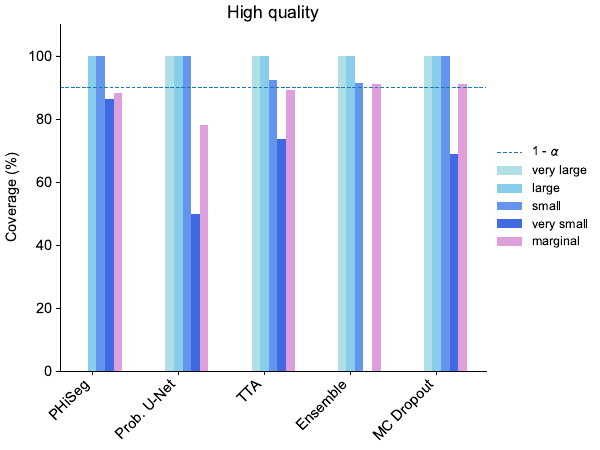}
    \includegraphics[width=0.49\textwidth]{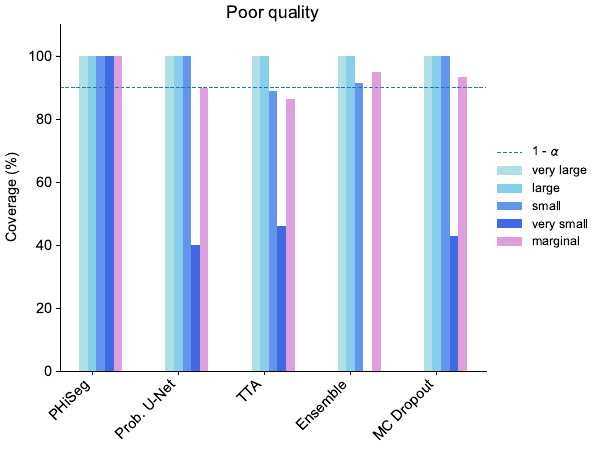}
    \caption{Marginal and conditional coverage for multiple interval sizes: very small (0, 0.1], small (0.1, 2], large (0.2, 5]), very large (0.5, 1]) for high quality images (left) and poor quality images (right).}  
\end{figure}

\end{document}